\documentclass[manuscript=article]{achemso}
\usepackage{amsmath,amsfonts,epsfig,graphicx,float}

\usepackage[version=3]{mhchem} 
\usepackage[T1]{fontenc}       

\title[]
  {Laser-Induced Spallation of Microsphere Monolayers 
   }

\author{Morgan Hiraiwa}
\author{Melicent Stossel}
\author{Amey Khanolkar}
\author{Junlan Wang}
\author{Nicholas Boechler}
\affiliation{Department of Mechanical Engineering, University of Washington, Seattle, WA 98195}

\email{boechler@uw.edu}

\begin{document}

\begin{abstract}
The detachment of a semi-ordered monolayer of polystyrene microspheres adhered to an aluminum-coated glass substrate is studied using a laser-induced spallation technique. The microsphere-substrate adhesion force is estimated from substrate surface displacement measurements obtained using optical interferometry, and a rigid-body model that accounts for the inertia of the microspheres. The estimated adhesion force is compared with estimates obtained from interferometric measurement of the out-of-plane microsphere contact resonance. Reasonable agreement is found between the two experiments. Scanning electron microscope images of detached monolayer regions reveal a unique morphology, namely, partially detached monolayer flakes composed of single hexagonal close packed crystalline domains. This work contributes to an improved understanding of microsphere adhesion and demonstrates a unique monolayer delamination morphology.  
 
\end{abstract}

\maketitle

\section{Introduction}

Microparticle adhesion is central to fundamental areas such as surface science \cite{IsraelachviliBook}, geological material mechanics \cite{Duran2000}, granular media dynamics \cite{3Resonance}, and even planet formation \cite{Dominik1997}. Improved understanding of microparticle adhesion has implications for applied areas including self-assembly \cite{SelfAssembly}, laser cleaning of semiconductors \cite{LaserCleaning}, powder processing \cite{Duran2000,Masuda2006}, and drug delivery \cite{DrugDeliveryReview}. One method to study microparticle adhesion is to use laser-generated acoustic waves to eject the particles from a surface. This method has previously been used to study the adhesion of disordered assemblies of particles adhered to substrates \cite{Maznev1998,Geldhauser2007}. A closely related technique, which is typically used to study interfacial adhesion of thin films, is laser-induced spallation \cite{Wang2002}, wherein, a laser-generated acoustic compression pulse reflects from the free surface of the sample as a tensile pulse that causes delamination near the surface. 

In this work, we study the delamination of semi-ordered microsphere monolayers adhered to an aluminum-coated glass substrate, which contains both hexagonal close packed (HCP) and disordered domains, using a laser-induced spallation technique. The particle-substrate adhesive force is estimated by monitoring the time-resolved displacements of the surface of a substrate without the microsphere monolayer via optical inteferometry, and applying a rigid-body model that accounts for the microsphere inertia. We compare the estimated adhesive force with estimations based on measurements of the out-of-plane contact resonance of the microspheres obtained using optical inteferometry \cite{3Resonance}. Scanning electron microscope (SEM) images of the delaminated monolayer areas reveal a unique morphology wherein the monolayer has fractured into flakes that have partially reattached to the substrate and are composed of mostly single crystalline domains.

\begin{figure}[H]
\begin{center}
\includegraphics[width=8.6 cm]{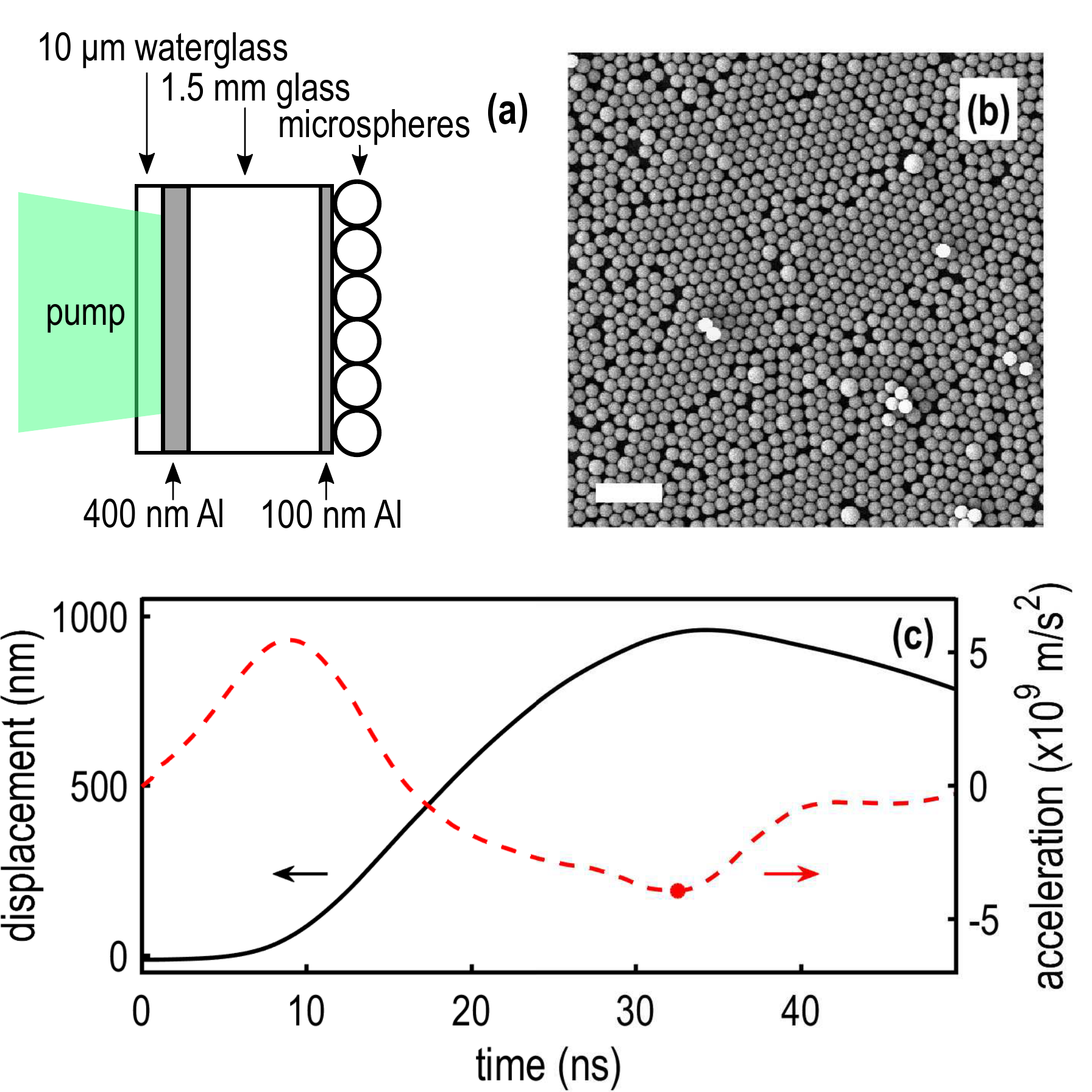}
\end{center}
\caption{\label{Figure1} (a) Schematic of the sample and the laser-induced spallation setup. (b) SEM image of an untested monolayer. The scale bar is 5 $\mu$m. (c) Measured surface displacement (black solid curve) and calculated surface acceleration (red dashed curve) of a substrate without a monolayer for a pump energy of 36 mJ. The marker indicates the identified point of maximum tensile force at the microsphere-substrate contact.}
\end{figure}  

\section{Experimental Section}
A schematic of the sample used in this study is shown in Fig. \ref{Figure1}a. The substrate was purchased from EMF Corp., and consists of a $1.5$ mm thick glass slide coated with a 100 nm thick aluminum film used to reflect light during the optical inteferometry measurements. A 400 nm thick aluminum film was applied on the opposite side via electron-beam evaporation in order to absorb the pump pulse laser light. A $10$ $\mu$m thick waterglass film is deposited on the back of the $400$ nm thick aluminum film via spincoating, to increase the amplitude of bulk longitudinal waves traveling through the substrate. A monolayer of $1$ $\mu$m diameter polystyrene (PS) microspheres are assembled on the top of the aluminum-coated substrate using a modified Langmuir-Blodgett technique \cite{Vogel}, in which the microspheres are self-assembled at an air-water interface and then transferred to the substrate. This process results in a semi-ordered monolayer, with HCP crystalline domains of differing orientations as well as highly disordered regions, as can be seen in Fig. \ref{Figure1}b. To excite acoustic waves in the substrate, pump laser light is focused on the $400$ nm aluminum layer. In all experiments, the plane of the sample surface is oriented vertically. 

In the laser-induced spallation experiments, the pump light (1064 nm wavelength, $5$ ns pulse duration) is focused to a spot size of approximately $2.5$ mm diameter (determined by burn marks on ZAP-IT paper). Since the excitation ablates the energy-absorbing aluminum layer, new monolayer positions are tested after each pump pulse. To calculate the force acting on the monolayer, surface displacements are monitored with a Michelson interferometer, which consists of an argon laser ($514.5$ nm wavelength, continuous wave) focused on the $100$ nm aluminum layer and the reference beam focused on a stationary mirror, to a spot with an estimated diameter of approximately $30$ $\mu$m (at 1/e\textsuperscript{2} intensity level). Because the microspheres scatter the probe light, surface displacements are recorded on a calibration sample that does not include a microsphere monolayer, but is otherwise identical. Interferometer signals are measured with a photodetector of $300$ ps rise time (Electro-Optics Technology, model ET-2030) and digitized on a 5 GHz oscilloscope (Lecroy Wavemaster 8500) at a sample rate of $20$ Gsample/s. Surface displacements are obtained from the interferometric signal using a fringe counting technique \cite{Wang2002}.

\section{Results and Discussion}

Figure \ref{Figure1}c shows an example of the measured surface displacements induced by a $36$ mJ pump pulse. We observe the arrival of an acoustic pulse of greater than $60$ ns duration traveling at the longitudinal sound speed in the glass. We obtain the acceleration of the substrate by numerically differentiating the measured substrate displacement twice, and applying numerical smoothing to the displacement, velocity, and acceleration signals \cite{Supp}. An example of such an acceleration profile, corresponding to a pump energy of $36$ mJ, is shown in Fig. \ref{Figure1}c. 

\begin{figure}[t]
\begin{center}
\includegraphics[width=8.6 cm]{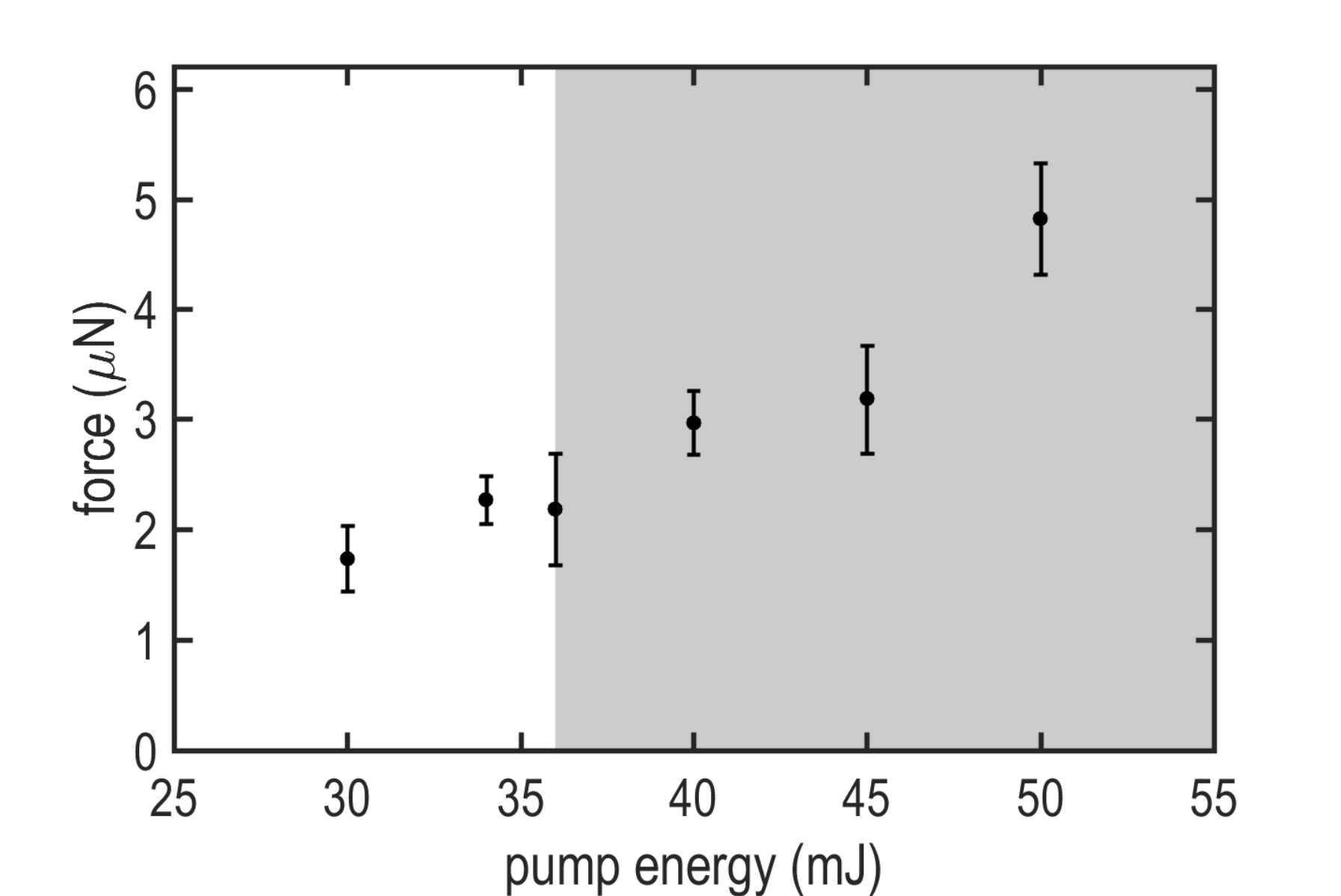}
\end{center}
\caption{\label{Figure2} Maximum tensile force induced at the microsphere contact as a function of laser pump energy. The shaded region indicates pump energies where microsphere detachment was observed.}
\end{figure}

To calculate the force acting at the particle-substrate contact, we employ a simple model that accounts for the inertia of the microspheres and considers the microspheres to be rigid bodies that follow the substrate surface motion until detachment \cite{Maznev1998}. We calculate the force applied to the contact as $F=ma$, where $m$ is the microsphere mass, calculated using a density provided by the manufacturer (Corpuscular, Inc.) of $1.06$ g/m\textsuperscript{3}, and $a$ is the measured substrate acceleration. This type of model is valid when the pulse duration is significantly longer than the period of the microsphere contact resonance (a vibrational mode where the microsphere moves like a rigid body, but has localized deformation around the point of contact that acts as a spring \cite{Audoin}). If the maximum tensile force induced at the contact exceeds the adhesive force, microsphere detachment occurs. The maximum tensile force obtained from the  acceleration profile is plotted against pump energy in Fig. \ref{Figure2}. The lowest pump energy where we observed microsphere detachment was at $36$ mJ, so we identify the detachment threshold to be between $36$ mJ and the next lowest pump energy, $34$ mJ. This corresponds to an adhesive force, averaged between $34$ mJ and $36$ mJ, of $F_{spall} = 2.2 \pm 0.4$ $\mu$N. The magnitude of the error bars in Fig. \ref{Figure2} is the maximum of the difference between the smoothed and unsmoothed acceleration signals, multiplied by the microsphere mass \cite{Supp}. As such, we note that the amplitude of the error bars do not account for inaccuracies that may result from fast changes in the substrate motion and the use of the inertial model, for instance, in the case of resonant particle removal where the rigid body assumption no longer holds.  

\begin{figure*}[t]
\begin{center}
\includegraphics[width=17 cm]{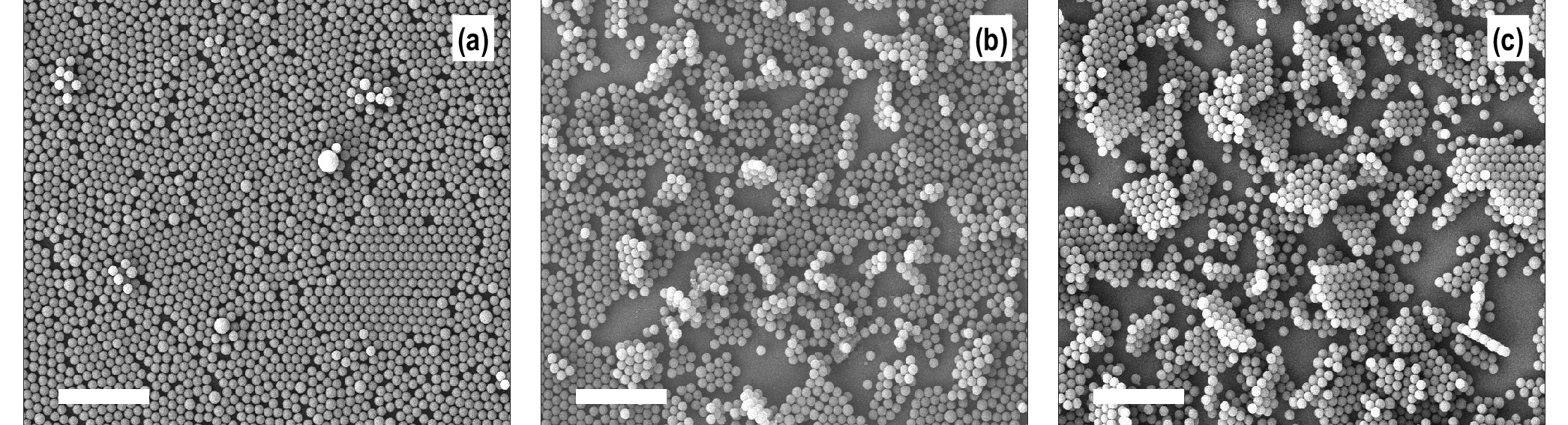}
\end{center}
\caption{\label{Figure3} SEM images of the monolayer after excitation with (a) $34$ mJ, (b) $36$ mJ, and (c) $50$ mJ pump pulses. The scale bars are 10 $\mu$m.}
\end{figure*}

After the spallation experiments, the sample is examined using a SEM. Images of the post-spallation monolayer are shown for different pump energies in Fig. \ref{Figure3}. A monolayer region tested with a pump energy of $34$ mJ is shown in Fig. \ref{Figure3}a, and appears similar to the untested region, with no evidence of microsphere detachment. When the pump energy is slightly increased to $36$ mJ, noticeable gaps in the monolayer are observed, as is shown in Fig. \ref{Figure3}b. At a pump energy of $50$ mJ, the amount of monolayer delamination increases, as can be seen in Fig. \ref{Figure3}c. Since the adhesive force is expected to have a statistical variation throughout the monolayer \cite{Blum1999}, it is not surprising that some patches of monolayer remain adhered to the surface near the onset of microsphere removal. We also observe that detached monolayer flakes resettle on the sample surface. Resettling of ejected particles has been observed previously by others \cite{Maznev1998}, despite the plane of the sample's surface being oriented vertically. This was explained by the slowing of ejected particles by the Stokes drag in air, while long range attractive forces eventually cause some particles to re-adhere to the surface \cite{Maznev1998}.  

We find that almost all of the delaminated flakes are composed of single HCP crystalline domains. We suggest that fracture of the monolayer into flakes is the result of interparticle decohesion along the weak boundaries that develop during the spallation process. In previous spallation experiments with continuous, ductile thin films, the excessive stress tended to lead to large scale plastic deformation in the film manifested by ``blister'' formation following the film delamination \cite{WangJMPS}. We speculate that the ordering of the monolayer flakes is due to stronger interparticle cohesion for HCP regions than disordered regions.

\begin{figure}
\begin{center}
\includegraphics[width=8.6 cm]{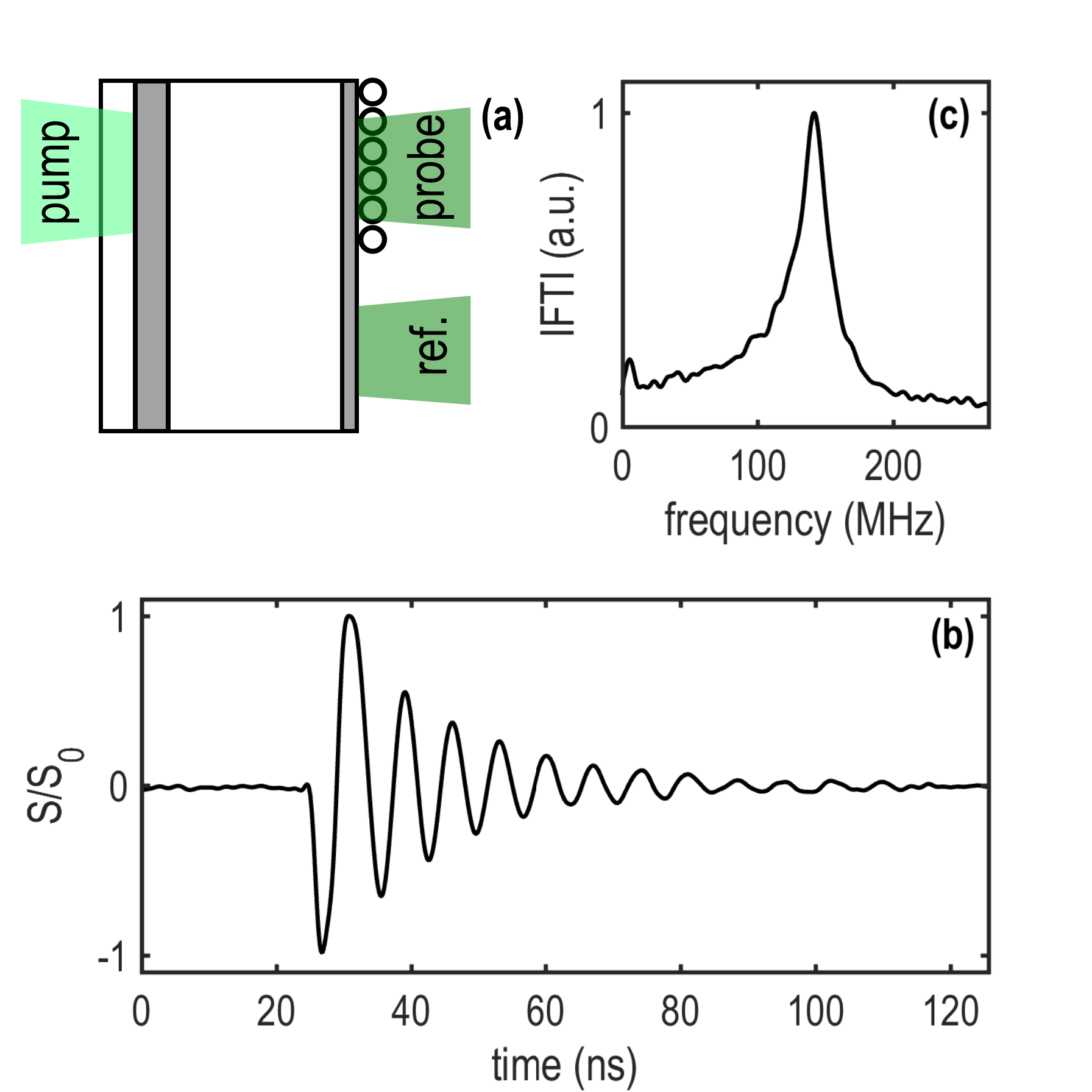}
\end{center}
\caption{\label{Figure4} (a) Schematic of the experimental setup used to measure the microsphere contact resonance. The sample is the same as in Fig. 1a. (b) Measured signal. (c) Fourier spectrum of (b).}
\end{figure}  

We compare the adhesion force obtained from the laser-induced spallation experiments with that estimated from the out-of-plane microsphere contact resonance measured on a different region of the same sample. To excite the contact resonance, we generate a bulk longitudinal acoustic pulse using a configuration similar to the spallation experiment, but at an amplitude significantly below the microsphere detachment threshold, as shown in Fig. \ref{Figure4}a. The out-of-plane contact resonance is excited upon arrival of the longitudinal wave, while the horizontal-rotational resonances \cite{Wallen2015} are not excited due to symmetry constraints. Out-of-plane displacements of the microspheres are measured using a grating interferometer \cite{Grating}, wherein a probe beam focused on the microspheres and a reference beam is focused on a stationary region of the sample. Probe and reference beams (514 nm wavelength, continuous wave) are both focused to a diameter of $ 80$ $\mu$m (at 1/e\textsuperscript{2} intensity level), while the pump (532 nm, 430 ps pulse duration, 4 $\mu$J per pulse) is focused to a diameter of $240$ $\mu$m (at 1/e\textsuperscript{2} intensity level). The interferometric signal is measured with a photodetector of $500$ ps rise time (Electro-Optics Technology, model ET-2030A), digitized with a 2.5 GHz oscilloscope (Tektronix DPO 7245C) at a sample rate of 5 Gsamples/s, and low-pass filtered with a cutoff frequency of 500 MHz. Using this setup, we obtain the signal shown in Fig. \ref{Figure4}b. The signal exhibits oscillations that begin when the bulk longitudinal acoustic wave arrives at the surface containing the microsphere monolayer. The Fourier spectrum of this signal is shown in Fig. \ref{Figure4}c, and has a clear peak at $f_{0}=140$ MHz. We relate the resonant frequency to the contact stiffness by assuming the spheres act as a spring mass oscillator with resonant frequency $f_{0}$, such that $f_{0}=1/2 \pi \sqrt{K_{c}/m}$. By linearizing the Hertzian contact model around the equilibrium contact area, we obtain the adhesive force in terms of the contact stiffness, $K_{c}$, such that $F_{CR} =(2 K_{c}/3)^{3}/RK^{2}$, where $F_{CR}$ is the adhesive force, $K$ is the effective modulus, defined as $K=[(3/4)(((1-\nu_{s}^{2})/E_{s})+((1-\nu_{l}^{2})/E_{l}))]^{-1}$, and R is the effective radius of curvature of the contact (equal to the microsphere radius for spherical-planar contact geometry) \cite{Hertz}. The effective modulus $K$ is expressed in terms of Young's modulus, $E$, and Poisson's ratio, $\nu$, and the subscripts $s$ and $l$ denote the sphere and the substrate, respectively. Using elastic properties of Young's modulus $E_{l}$ = 62 GPa and Poisson's ratio $\nu_{l}$ = 0.24 for the aluminum layer \cite{ASMHandbook}, and $E_{s}$ = 4.04 GPa and $\nu_{s}$ = 0.32 for the PS microspheres \cite{Khanolkar2015}, we find an adhesive force of $F_{CR}= 1.5$ $\mu$N.     

Using the Derjaguin-Muller-Toporov model for adhered microparticles \cite{DMT,FurtherDMT}, we calculate an adhesive force of $F_{DMT} = 2 \pi wR  = 0.36$ $\mu$N, where $w = 0.113$ J/m\textsuperscript{2} is work of adhesion between PS and alumina (assuming a native oxide layer) \cite{Khanolkar2015}. While the adhesion force obtained via the two experiments have reasonable agreement, the adhesion estimated using DMT contact mechanics is significantly lower. Previous studies have similarly found larger than predicted contact stiffnesses \cite{ 3Resonance,Khanolkar2015}. Uncertainties in the work of adhesion between the PS microparticle and the aluminum-coated substrate may contribute to the higher than predicted adhesion force. Particularly in the case of reactive metals, such as aluminum, higher values of work of adhesion have been observed than are predicted by van der Waals adhesion models \cite{Israelachvili}. Utilizing the DMT contact model and the adhesion force obtained from the laser-induced spallation experiments, we obtain a work of adhesion of $0.8$ J/m\textsuperscript{2}, which is reasonable considering past studies on dielectric-metal adhesion \cite{Israelachvili}. By observing PS microspheres adhered to substrates via SEM, previous studies have found the contact radius to be larger than predicted by elastic contact models \cite{Rimai1989,Rimai1990}, which can also result in higher adhesive forces \cite{ParticlesBook,PlasticContact}.

\section{Summary and Conclusions}
This work provides data regarding the adhesion, contact mechanics, and delamination of microsphere monolayers. The unique monolayer flakes may find future application in laser-induced particle transfer applications \cite{Kattamis2007}, and provide an opportunity to study the in-plane fracture mechanics of microsphere monolayers. The microsphere monolayer studied in this work represents a microscale analog of a macroscale granular crystal \cite{BoechlerBook}. While many studies have shown macroscale granular crystals to exhibit rich physics, the study of their microscale counterparts is an emerging field \cite{PhysicsToday}. This work paves the way for the study of spallation and highly nonlinear phenomena in three-dimensional microscale granular crystals. Related laser-induced spallation techniques generating shear \cite{WangPureShear} or mixed-mode \cite{WangJMPS} elastic waves may be useful to study the shear strength of adhered microparticles, or the propagation of rotational modes in three dimensional microscale granular crystals. Rotational modes have only recently been observed in three-dimensional macroscale \cite{Merkel2011} and two-dimensional microscale granular crystals \cite{3Resonance}. Finally, the microparticle adhesion measurements obtained as part of this study may find future use in self-assembly, laser cleaning, powder processing, or drug delivery applications. 

%
\section{Acknowledgements}

The authors appreciate useful discussions with S. P. Wallen, N. Vogel, and A. A. Maznev. N.B. and A.K. acknowledge funding from the US Army Research Office (grant no. W911NF-15-1-0030) and the US National Science Foundation (grant no. CMMI-1333858). M.H. acknowledges support from the National Science Foundation Graduate Research Fellowship Program under Grant No. DGE-1256082. J.W. and M.S. acknowledge financial support from the University of Washington Clean Energy Institute.

\section{Supporting Information}

The supporting information contains additional details concerning the data processing procedure.
 
%
%

%
%
\end{document}